\documentclass[conference]{IEEEtran}

\usepackage{cite}
\usepackage{color}

\usepackage[pdftex]{graphicx}
\graphicspath{{../img/}}

\usepackage{booktabs}

\usepackage[cmex10]{amsmath}
\usepackage{array}
\usepackage{mdwmath}
\usepackage[caption=false,font=footnotesize]{subfig}
\usepackage{url}
\usepackage[keeplastbox]{flushend}
\usepackage{listings}
\usepackage[usenames,dvipsnames,svgnames,table]{xcolor}
\usepackage{courier}

\usepackage[hidelinks]{hyperref}

\usepackage[caption=false,font=footnotesize]{subfig}

\usepackage[super]{nth}

\usepackage{array}
\usepackage{multirow}
\usepackage{multicol}

\usepackage{algorithm}
\usepackage[noend]{algpseudocode}

\algrenewcommand{\alglinenumber}[1]{\scriptsize #1:}
\algrenewcommand{\algorithmicindent}{0.5em}		
\algrenewcommand{\algorithmiccomment}[1]{$//$ #1}	

\algblockdefx[subProc]{START}{END}%
             [2]{#1(#2)}%
             [1]{\textbf{end} #1}

\algblockdefx[subProc]{ForEach}{EndFor}%
             [1]{\textbf{for each} #1 \textbf{do}}%
             {\textbf{end for}}

\algtext*{EndFor}	

\makeatletter
\renewcommand\fs@ruled{%
  \def\@fs@cfont{\rmfamily}%
  \let\@fs@capt\floatc@plain%
  \def\@fs@pre{}
  \def\@fs@post{}
  \def\@fs@mid{}
  \let\@fs@iftopcapt\iffalse}
\makeatother

\begin{document}

\title{A Scalable Approach for Hardware\\Semiformal Verification}

\author{
  \IEEEauthorblockN{Tom\'as Grimm\IEEEauthorrefmark{1}, Djones Lettnin\IEEEauthorrefmark{2}, Michael H\"ubner\IEEEauthorrefmark{1}}
  \IEEEauthorblockA{
    \IEEEauthorrefmark{1}Chair of Embedded Systems for Information Technology (ESIT), Ruhr-University Bochum, Germany\\
    \IEEEauthorrefmark{2}Department of Electrical and Electronics Engineering, Federal University of Santa Catarina, Brazil\\
    Email: Tomas.Grimm@rub.de, djones.lettnin@ufsc.br, Michael.Huebner@rub.de
  }
}


\maketitle

\begin{abstract}
The current verification flow of complex systems uses different engines synergistically: virtual prototyping, formal verification, simulation, emulation and FPGA prototyping. However, none is able to verify a complete architecture. Furthermore, hybrid approaches aiming at complete verification use techniques that lower the overall complexity by increasing the abstraction level. This work focuses on the verification of complex systems at the RT level to handle the hardware peculiarities. Our results show an improvement of 100\% compared to the commercial tool's results for the prototype we used to validate our approach.
\end{abstract}

\section{Introduction}
\label{sec:introduction}
Systems-on-chip (SoCs) are widespread nowadays, covering a wide spectrum of electronics, e.g.\ in cell phones, tablets, and cars. This variety of applications means that SoCs' complexity is increasing and will keep increasing in the next generations~\cite{ITRS2015}. The ITRS System Integration group predicts, for a single SoC architecture, an increase in application processors from 9 elements in 2017 to 18 in 2020 and 36 in 2025 and in graphics processing units from 19 elements in 2017 to 58 in 2020 and 247 in 2025. Still according to~\cite{ITRS2015}, ``the degree of integration after 2008 keeps increasing to meet the demands of (i) higher computation performance, (ii) faster wireless connections, and (iii) richer multimedia capabilities.''

However, the increase in complexity and functionality has a hidden cost: ``The increasing number of heterogeneous components (RF, logic, memory, and MEMS) complicates the system design''~\cite{ITRS2015} and complete verification of such systems is practically impossible~\cite{Turley2013}. Different engines exist to try to solve the verification problem at each stage of development. Simulation, emulation, FPGA prototyping, and formal verification are currently the preferred engines for hardware verification in the industry~\cite{Foster2015}. Nevertheless, so far no engine is capable of giving 100\% coverage for a complex architecture.

The research field for solutions to the verification gaps is wide and strong both in the academia and in the industry and different techniques and methodologies exist~\cite{Singhal}. For example, there are techniques to either verify an architecture in a higher level of abstraction~\cite{Cimatti2013} or to divide it into smaller sub-blocks and verify them separately. Both have advantages and disadvantages. The first technique speeds up the verification process by increasing the abstraction, but it gives meaningful results only for the functionality of the hardware, not for its low-level behavior, e.g.\ timing and parallelism. Therefore, it is useful only for the initial phases of development. The second technique lowers the system complexity by verifying small portions instead of the whole system at once; however, it is unable to cover the interaction between the many sub-blocks that compose the system.

Even though these and many other techniques have their advantages, it is still necessary to verify the entire architecture in the later stages of development before moving to the physical implementation. It is only by stimulating the complete architecture that all its functionalities can run in parallel and highlight the corner cases that need deeper verification.

Coming back to the four engines used in the industry, each serves a different purpose and applies to different stages of development.

Formal verification can completely verify IPs and small subsystems, allowing for complete coverage~\cite{seligman2015formal} and fits well in the early stages of development. Nevertheless, it does not scale well for architectures that are more complex due to state space explosion.

Simulation and emulation stimulate an architecture with specific test vectors to generate intermediate and output values~\cite{Gajski2009} and they fit best later in the development flow when the architecture is more stable and all sub-blocks are verified. Although both are scalable, they cannot cover every possible test case.

FPGA prototyping applies to mature architectures~\cite{Nenni2016} to allow for at-speed testing with the embedded software, however, at the expense of reduced internal observability.

To improve the verification results for big architectures, a new trend currently becoming more popular in the industry is the synergy between verification engines~\cite{Simm2015}, where the verification team seeks to combine the advantages of each engine to the most applicable development phase. One simple but powerful example is deep dynamic formal verification~\cite{EDA-013}. Simulation directs the architecture to a specific state, and from there, formal tools try to verify a smaller set of states. This technique relies on the quality of the input vectors to drive the system to the desired deep space or corner case. However, this can be challenging and time-consuming to perform iteratively, mostly due to the need to simulate millions of cycles to reach the desired state.

The aim of this work is to increase coverage using dynamic data to cover a greater set of states, without resorting to deep states. As SoC architectures grow in complexity in every new generation of electronics, this growth highlights the need for new approaches to the verification problem. This work proposes a ``build-and-prove'' process tied to a static register assignment (SRA) heuristic to reduce the state space for formal verification.

The main contribution of this work is a scalable, hybrid, iterative and embedded software simulation-driven flow to improve the verification productivity. The build-and-prove process tries to verify subsystems that grow at each iteration until it reaches the complete architecture. Simulation runs help to reduce the state space for the formal verification process and to avoid the state space explosion. The simulation process uses the embedded software to provide the dynamic data to constrain the architecture during the semiformal phases. The SRA process tries to improve the constraints in an iterative fashion.

The next sections of this paper are organized as follows: Section~\ref{sec:related} describes the related work. Section~\ref{sec:content} presents the developed work in detail. Section~\ref{sec:results} summarizes the results after applying this work to a test case. Finally, Section~\ref{sec:conclusion} concludes this paper.

\section{Related Work}
\label{sec:related}
Mukherjee et al.~\cite{Mukherjee2015} propose a flow to translate RTL code to ANSI-C code and apply different formal techniques for software to it, e.g.\ bounded model checking, path-based symbolic simulation, and abstract interpretation. The idea is to increase the abstraction level and simplify the proofs in order to get results faster. However, due to hardware's nature, software models cannot accurately describe some of its peculiarities. An example is the generation of netlists from high-level models using technologies such as high-level synthesis (HLS). Since current HLS tools cannot capture specific hardware details from the software description, e.g.\ parallelism and pipelining, they do not implement the developer's intent correctly.

Herber~\cite{Herber2014} aims at hardware/software co-verification by partitioning SystemC models to achieve a scalable flow. In this work, different engines verify different aspects of an architecture. For instance, a satisfiability-modulo-theory (SMT) tool verifies synchronous components of the RTL. The proposed tool splits the verification task into hardware, software, and system level. However, as useful as this approach may be, it is only applicable to the initial stages of design since the granularity level is too coarse for deep verification.

Gro{\ss}e et al.~\cite{Groe2006} divide the formal verification of SystemC models into three steps. The first step checks the hardware blocks separately. The second step does the verification of the hardware/software interface using the results from the previous step. The third step verifies the embedded software (ESW). As it is the case with the previously cited works, this approach targets a higher-level description, i.e. SystemC, and thus fails to address characteristics inherent to the hardware, focusing only on its functionalities.

The above-mentioned works show a gap in low-level verification since all approaches employ higher-level abstractions to try to improve verification results at the expense of fine-grained details. We close this gap with this work by only focusing on the RT-level and using a ``build-and-prove'' system tied to a novel heuristic to avoid the state-space explosion as much as possible.

\section{Scalable Semiformal Hardware\\Verification Methodology}
\label{sec:content}
As aforementioned, large systems cannot be completely verified using formal methods due to state space explosion. To overcome this problem, we developed an iterative build-and-prove system. It starts the verification process proving a small subsystem and increasing it iteratively IP-by-IP. The verification begins with formal methods and, when they become insufficient, a proposed semiformal heuristic aids to overcome the system complexity.

We propose the HWVerifyr verification approach in this work, which has five phases: (1) RTL preprocessing, (2) formal and (3) semiformal verification of the IPs, (4) formal and (5) semiformal verification of subsystems using the build-and-prove process. Algorithm~\ref{alg:semiformal_flow} describes the proposed flow.

The next sections describe each phase in detail as well as our developed SRA heuristic to improve the verification process.

\begin{algorithm*}[t]
  \small
  \caption{The HWVerifyr}
  \label{alg:semiformal_flow}

  \begin{multicols}{2}
    \begin{algorithmic}[1]
      \START{HWVerifyr}{ESW, PropSet, RTL}
        \START{doPreprocessing}{}						\label{startPreprocessing}
          \State{IPs = listUniqueIPs(RTL)}					\label{phase1_listIPs}
          \State{rankedIPs = rankIPsByConnection(RTL)}				\label{phase1_rankedIPs}
          \State{mergedESW = cilly(ESW)}					\label{phase1_cilly}
          \State{mRegs = mapRegs(RTL, mergedESW)}				\label{phase1_mappedRegs}
          \State{propGroups = divideProps(PropSet)}				\label{phase1_groups}
        \END{doPreprocessing}							\label{endPreprocessing}

        \Statex{}

        \START{doIPsVerification}{}						\label{startFormalIPs}
          \ForEach{IP in IPs}							\label{phase2_loop}
            \State{s = startBMC(IP, PropSet)}					\label{phase2_ipBMC}
            \If{s = false}
              \State{marked += IP}                                              \label{phase2_addIPtoList}
            \EndIf{}
          \EndFor{}                                           			\label{endFormalIPs}

          \Statex{}

          \If{!Empty(marked)} \label{startSemiformalIPs}
            \State{rRegs = doSRA(marked)}					\label{phase3_sra}             
            \State{POIs = setPOIs(mRegs, rRegs)}                                \label{phase3_pois}            
            \State{startSimulation()}                                           \label{phase3_sim}             
            \While{!Empty(marked)}                                              \label{phase3_poi_loop}        
              \If{watchTriggeredPOItriggered(POIs)}                             \label{phase3_poi_triggered}   
                \While{!Empty(rRegs)}                                           \label{phase3_reg_loop}        
                  \State{simVals = collectSimValues()}                          \label{phase3_sim_vals}        
                  \State{stopats = createStopats(rRegs)}                        \label{phase3_stopats}         
                  \State{asm = createAssumes(simVals)}                          \label{phase3_assumes}         
                  \State{s = startBMC(IP, PropSet, stopats, asm)}               \label{phase3_bmc}     
                  \If{s = false}                                                                               
                    \If{Empty(rRegs)}                                           \label{phase3_test}            
                      \If{blackboxFailingIPs = true}                                                           
                        \State{blackbox(IP)}                                    \label{phase3_bbox}            
                      \Else{}                                                                                  
                        \State{\textbf{return} SEMIFORMAL\_FAIL}                \label{phase3_return}          
                      \EndIf{}                                                                                 
                    \Else{}                                                                                    
                      \State{combineRegs(rRegs)}                                \label{phase3_combine_regs}    
                    \EndIf{}                                                                                   
                  \Else{}                                                                                      
                    \State{marked -= IP}                                        \label{phase3_removeIP}        
                  \EndIf{}
                \EndWhile{}
              \EndIf{}
            \EndWhile{}
          \EndIf{}								\label{endSemiformalIPs}
        \END{doIPsVerification}



        \START{doSubsystemsVerification}{}					\label{startFormalSubsys}
          \State{subsys = createSubsystem(rankedIPs)}				\label{phase4_create}
          \While{!Empty(rankedIPs)}                                             \label{phase4_loop}
            \State{s = startBMC(subsys, PropSet)}                               \label{phase4_bmc}
            \If{s = false}
              \State{switchToSemiformal()}                                      \label{phase4_switch}
            \Else{}
              \If{!Empty(rankedIPs)}
                \State{subsys = addIPtoSubsystem(rankedIPs)}                    \label{phase4_addIP}
              \Else{}
                \State{\textbf{return} FORMAL\_COMPLETE}                        \label{phase4_return}
              \EndIf{}
            \EndIf{}
          \EndWhile{}                                                           \label{endFormalSubsys}

          \Statex{}

          \While{!Empty(rankedIPs)}						\label{startSemiformalSubsys}
            \State{rRegs = doSRA(subsys)}					\label{phase5_sra}                  
            \State{POIs = setPOIs(rRegs)}                                       \label{phase5_pois}                 
            \State{startSimulation()}                                           \label{phase5_sim}                  
            \If{POItriggered(POIs)}                                             \label{phase5_poi_triggered}        
              \While{!Empty(rRegs)}                                             \label{phase5_reg_loop}             
                \State{simVals = collectSimValues()}                            \label{phase5_sim_vals}             
                \State{stopats = createStopats(rRegs)}                          \label{phase5_stopats}              
                \State{asm = createAssumptions(simVals)}                        \label{phase5_assumes}              
                \State{s = startBMC(subsys, PropSet, stopats, asm)}             \label{phase5_bmc}  
                \If{s = false}                                                                                      
                  \If{Empty(rRegs)}                                             \label{phase5_test}                 
                    \State{\textbf{return} SEMIFORMAL\_FAIL}                    \label{phase5_return_fail}          
                  \Else{}                                                                                           
                    \State{combineRegs()}                                       \label{phase5_combine_regs}         
                  \EndIf{}                                                                                          
                \Else{}                                                                                             
                  \If{Empty(rankedIPs)}                                                                             
                    \State{\textbf{return} SEMIFORMAL\_COMPLETE}                \label{phase5_return_success}       
                  \Else{}                                                                                           
                    \State{subsys = addIPtoSubsystem(rankedIPs)}                \label{phase5_build}
                  \EndIf{}
                \EndIf{}
              \EndWhile{}
            \EndIf{}
          \EndWhile{}
        \END{doSubsystemsVerification}						\label{endSemiformalSubsys}

        \Statex{}

        \State{\textbf{output} coverage}
      \END{HWVerifyr}
    \end{algorithmic}
  \end{multicols}
\end{algorithm*}

\subsection{Static Register Assignment Heuristic}
\label{sra}
The Static Register Assignment (SRA) heuristic's goal is to reduce the state space using dynamic information. Each time HWVerifyr calls SRA, it uses information from the simulation run.

To achieve the best results, it is important to scale down the state space without over-constraining it; otherwise, the constraints can make errors unreachable. SRA addresses this point using a register mapping between RTL and embedded software, which are elements reachable from the ``user'' side. This avoids using elements that the user has no control over, e.g. I/O interfaces.

SRA begins by building the DUV's netlist, either a single IP or a subsystem. From this netlist, SRA calculates the cone-of-relevance (COR) for each of the mapped registers and ranks them from highest to lowest. The output is a list of ranked registers for the semiformal verification phases.

\begin{figure}
  \centering
  \includegraphics[width=\columnwidth]{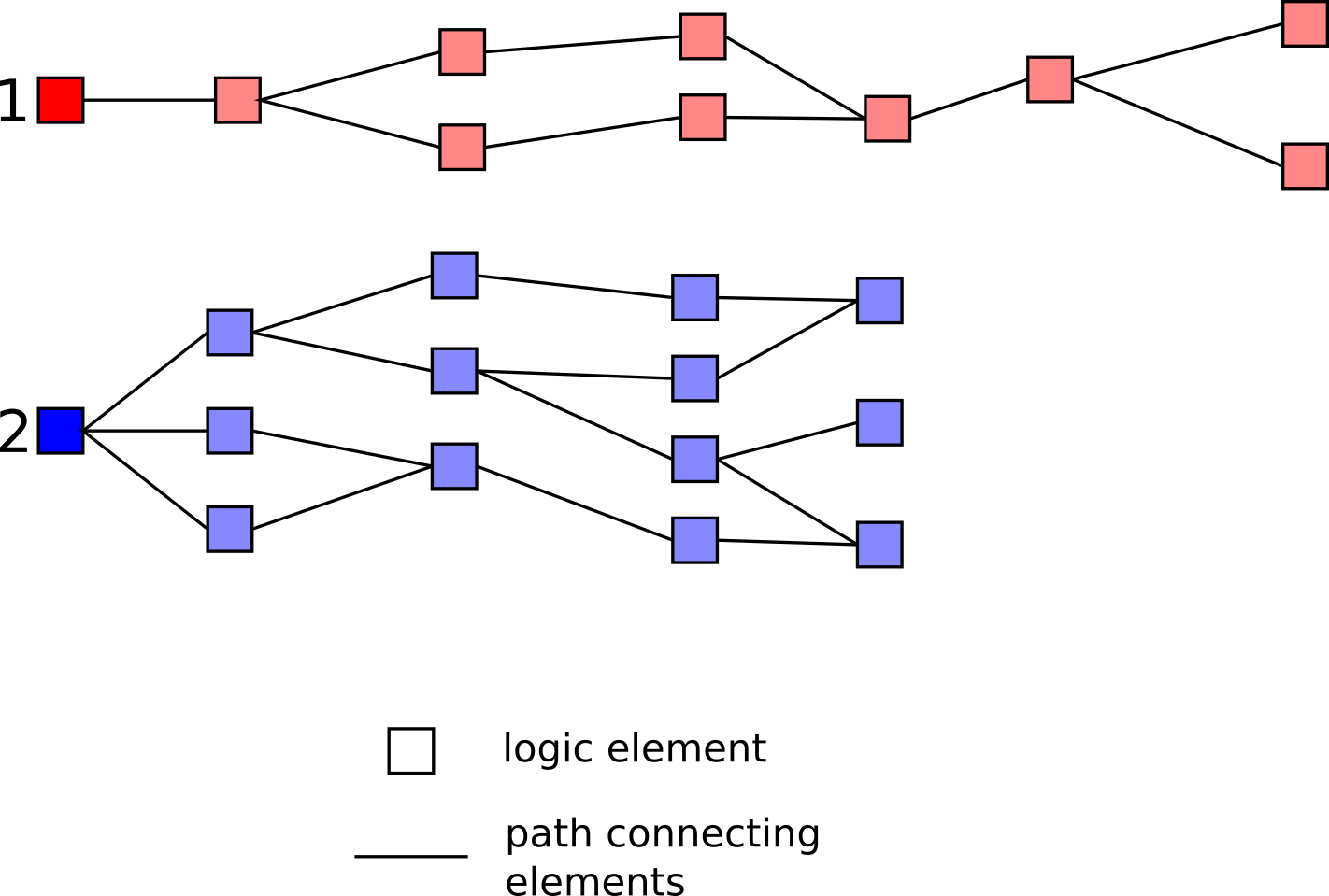}
  \caption{Illustration of the COR metric.}
  \label{fig:cor}
\end{figure}

The COR is a measure for the influence of a register based on its breadth and depth. The breadth indicates how many paths start at the register. The depth indicates how many state elements a register connects to, either directly or indirectly. Breadth receives a greater weight due to the register's influence on multiple paths. The measure starts at each mapped register and covers all the logic from them to the outputs. For each path connected to a register, the COR for that register increases by 100 points and for each element connected along each path it increases by 1 point. Figure~\ref{fig:cor} presents a graphical illustration of this concept.

In Figure~\ref{fig:cor}, register 1 connects to one path and nine logic elements and register 2 connects to 3 paths and 13 logic elements. The resulting list has register 2 in first place and register 1 in second, since register 1 has a deeper but narrower relevance and register 2 has a shorter but broader relevance.

\subsection{Phase 1: Preprocessing}
Phase 1 (Algorithm~\ref{alg:semiformal_flow}, lines~\ref{startPreprocessing}--\ref{endPreprocessing}) is a preparation to the next phases. It prepares some of the data structures required later on. Phase 1 has five actions.

First, it extracts, from the RTL, the IPs that compose the architecture (Line~\ref{phase1_listIPs}). The result is a list of all unique IPs in the architecture used by phases 2 and 3. These phases do the verification of each IP separately.

Second, it ranks the instantiated IPs according to their connectivity (Line~\ref{phase1_rankedIPs}). The result is the ranked list of elements used by phases 4 and 5. These phases are the base of the iterative build-and-prove process introduced in Section~\ref{phase4}.

Third, it merges all source files to generate a single file, which has all accessed addresses explicitly encoded. We use the tool cilly (Line~\ref{phase1_cilly}), from the C Intermediate Language framework [9], to perform this action.

Fourth, it maps the registers between ESW and RTL using the single source file generated in the previous action (Line~\ref{phase1_mappedRegs}). From this file, HWVerifyr generates the mapping between the implementation of the RTL registers and the locations where the ESW accesses them. This mapping is the key point of the SRA heuristic developed in this work.

Fifth, it groups the user-provided formal properties by the IP(s) they cover (Line~\ref{phase1_groups}). The model checker uses these groups in all following phases.

\subsection{Phase 2: Formal verification of the IPs}
Phase 2 (Algorithm~\ref{alg:semiformal_flow}, lines~\ref{startFormalIPs}--\ref{endFormalIPs}) tries to verify all IPs separately with the model checker (Line~\ref{phase2_ipBMC}) to find any errors before starting the subsystems phase. It adds all the IPs that do not successfully complete within the time limit to a list for Phase 3 (Line~\ref{phase2_addIPtoList}). The default time limit to verify each IP is 3600 seconds, but the user can set a different value through a parameter.

\subsection{Phase 3: Semiformal verification of the IPs}
HWVerifyr executes Phase 3 (Algorithm~\ref{alg:semiformal_flow}, lines~\ref{startSemiformalIPs}--\ref{endSemiformalIPs}) if there are any IPs in the output list of Phase 2. Otherwise, the tool goes to Phase 4.

Phase 3 begins by running the SRA process to generate the list of ranked registers for each IP marked for semiformal verification (Line~\ref{phase3_sra}). Next, it sets up the points-of-interest (PoIs) for the simulation (Line~\ref{phase3_pois}), which are the locations in the ESW that access any of the mapped registers. Following this, the simulation starts (Line~\ref{phase3_sim}). Whenever the simulator executes an instruction that involves a PoI (Line~\ref{phase3_poi_triggered}), the semiformal verification process begins. First, HWVerifyr communicates with the simulation engine to collect the dynamic data for the current ranked registers (Line~\ref{phase3_sim_vals}). Next, it creates stop-ats associated with the registers (Line~\ref{phase3_stopats}) and uses the values from the simulation to generate assumptions for the registers' outputs (Line~\ref{phase3_assumes}). The formal tool, then, tries to prove the properties associated with the current IP using the generated stop-ats and assumptions (Line~\ref{phase3_bmc}). The verification runs for the duration of the time limit. If it cannot complete and there are no more registers left, the tool either black boxes the IP (Line~\ref{phase3_bbox}), if the user chose to black box failing IPs or it aborts the process and outputs the results to the user (Line~\ref{phase3_return}). If the list of ranked registers still has elements, the tool uses the next in the sequence and restarts the semiformal verification process (Line~\ref{phase3_combine_regs}). If the model checker verifies the IP successfully, the tool removes it from the list and resumes the simulation (Line~\ref{phase3_removeIP}). This loop executes until all IPs are either verified or black-boxed or if it fails.

A stopat is an abstraction used to ``cut'' the driving logic beyond a chosen point. This enables the model checker to choose a value for a proof. Furthermore, assumptions can tell the model checker which value it must use from that point on.

Black boxing instructs the formal tool to ignore the internal architecture for some block and unconstrain all its output signals.

\subsection{Phase 4: Formal verification of the subsystems}
\label{phase4}
After HWVerifyr verifies every IP successfully, it starts the build-and-prove iterative proving process. This process uses the list of ranked IPs from the preprocessing phase to build subsystems that grow by one IP at each iteration. It begins the subsystem with the two highest ranked IPs, verifies it and, after success in the verification, adds the next highest ranked IP from the list and repeats the process. The build process follows the original architecture.

Phase 4 (Algorithm~\ref{alg:semiformal_flow}, lines~\ref{startFormalSubsys}--\ref{endFormalSubsys}) begins building the first subsystem (Line~\ref{phase4_create}) and verifying it (Line~\ref{phase4_bmc}). If the model checker's status is incomplete, HWVerifyr switches to Phase 5 for the semiformal approach (Line~\ref{phase4_switch}). Otherwise, if the proof is successful and the list of IPs is not yet empty, the tool adds the next IP to the subsystem (Line~\ref{phase4_addIP}) and restarts the verification process. If the list is empty, then the process ends successfully (Line~\ref{phase4_return}).

\subsection{Phase 5: Semiformal verification of the subsystems}
Phase 5 (Algorithm~\ref{alg:semiformal_flow}, lines~\ref{startSemiformalSubsys}--\ref{endSemiformalSubsys}) continues the iterative build-and-prove process using the SRA heuristic. It starts from the unsuccessful subsystem from Phase 4.

As in Phase 3, Phase 5 begins by ranking the subsystem's registers (Line~\ref{phase5_sra}), setting the PoIs for the simulation (Line~\ref{phase5_pois}) and triggering the simulation (Line~\ref{phase5_sim}). When the simulator executes an instruction with a PoI (Line~\ref{phase5_poi_triggered}), the semiformal verification starts with the highest ranked register. HWVerifyr collects the values from the simulation (Line~\ref{phase5_sim_vals}), adds the corresponding stop-at for the register (Line~\ref{phase5_stopats}), generates the necessary assumptions from the simulation data (Line~\ref{phase5_assumes}) and calls the model checker (Line~\ref{phase5_bmc}). If the verification status is incomplete and the list of ranked registers is empty, the process aborts (Line~\ref{phase5_return_fail}), otherwise, HWVerifyr adds the next ranked register and restarts the semiformal verification process (Line~\ref{phase5_combine_regs}). If the verification status is complete and the list of IPs is empty, the tool returns success (Line\ref{phase5_return_success}); otherwise, it adds the next IP to the subsystem and restarts the phase with the updated subsystem (Line~\ref{phase5_build}).

The starting set of registers for this phase contains all the registers that were successful at the end of Phase 3.

\section{Results and Discussion}
\label{sec:results}

\subsection{Verification Environment}
We used a 24 core Intel\textregistered\ Xeon\textregistered\ CPU E5--2630 @ 2.3GHz with 96GB of RAM memory running CentOS to perform the experiments. We used the tool JasperGold to obtain the baseline results and we compared these results to those obtained using our scalable hybrid approach described in Section 3. We also used JasperGold as the model checker for our experiments.

Our experiments used the X-Propagation app from JasperGold to extract the properties for each architecture, either IP or subsystem, and verify them. Table~\ref{tab:properties} shows the number of properties extracted for each IP and for each subsystem.

\begin{table}%
  \caption{Number of X-Propagation Properties Generated for Each Architecture}
  \begin{minipage}{\columnwidth}
    \begin{center}
      \begin{tabular}{lclc}
        \toprule
        \multicolumn{2}{c}{IPs} & \multicolumn{2}{c}{Subsystems} \\
        \midrule
        Architecture & \# props & Architecture & \# props        \\
        \midrule                                
        CAN          & 1274     & Subsystem 1  & 2364            \\
        ETHMAC       & 3207     & Subsystem 2  & 3643            \\
        MOR1KX       & 1644     & Subsystem 3  & 6847            \\
        WB\_RAM      & 93       &              &                 \\
        \bottomrule
      \end{tabular}
    \end{center}
  \end{minipage}
  \label{tab:properties}
\end{table}%

\subsection{Automotive Gateway Prototype}
\begin{figure}
  \centering
  \includegraphics[width=\columnwidth]{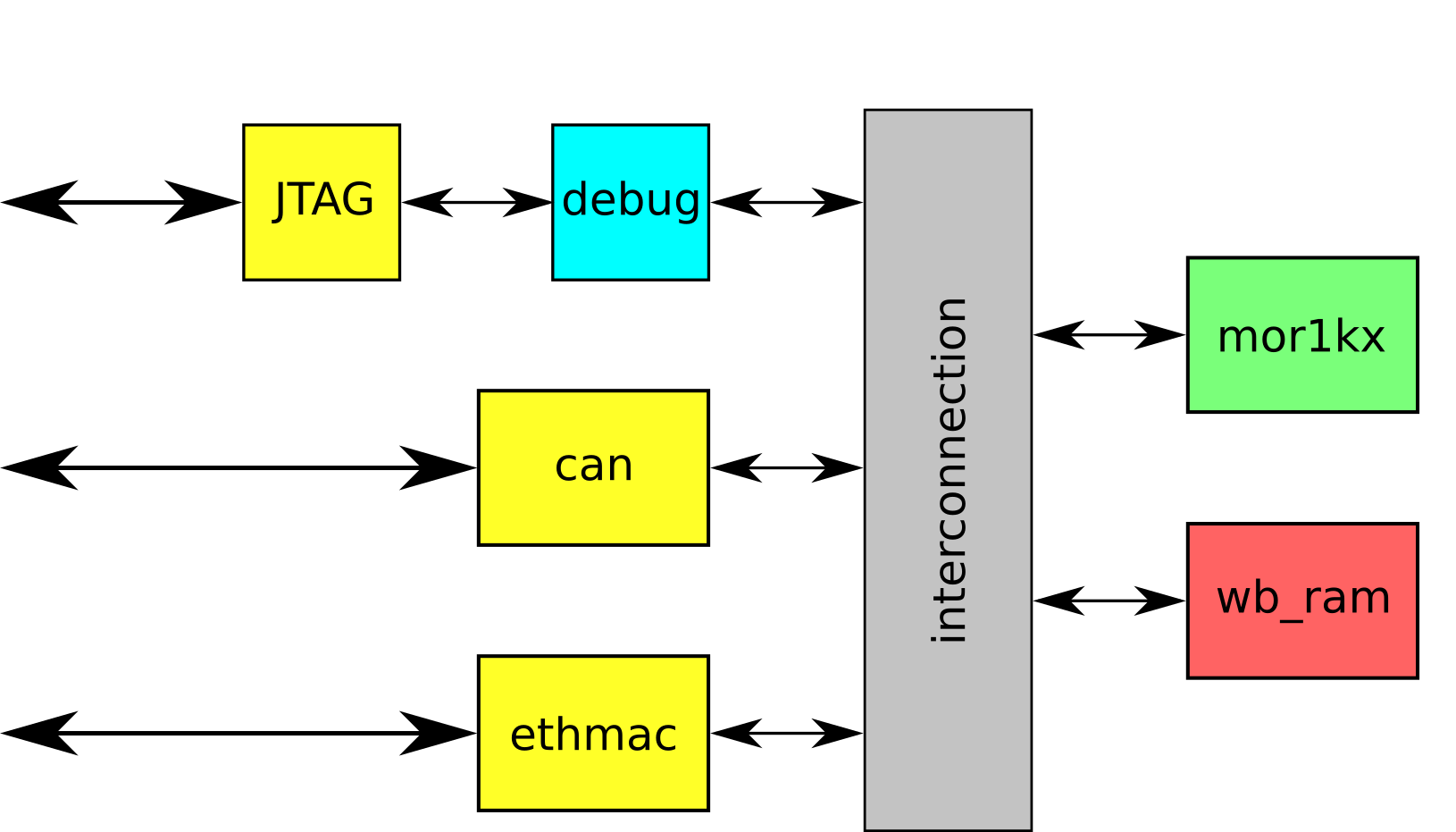}
  \caption{Prototype's block diagram as generated by the Fusesoc platform.}
  \label{fig:gateway}
\end{figure}

We used a prototype for an automotive gateway developed in-house as our case study. We develop this prototype using the Fusesoc platform. It has an OpenRISC processor (``MOR1KX''), a CAN IP (``CAN''), an Ethernet IP (``ETHMAC''), and a RAM memory IP (``WB\_RAM''). All elements are from the OpenCores repository. Figure~\ref{fig:gateway} presents its architectural block-level diagram and Table~\ref{tab:comparison} summarizes the results for the validation of HWVerifyr. The chosen time limit for each IP was 3600 seconds and for each subsystem was 5400 seconds.

\begin{table}%
  \caption{Verification Results for the Gateway Prototype}
  \begin{minipage}{\columnwidth}
    \begin{center}
      \begin{tabular}{lclcl}
        \toprule
        \multirow{2}{*}{Architectures} & \multicolumn{2}{c}{JasperGold}             & \multicolumn{2}{c}{Hybrid}                \\
        \cmidrule(lr){2-3}
        \cmidrule(lr){4-5}
                                       & Result           & Outcome                 & Result   & Outcome                        \\
        \midrule
        CAN                            & Timeout          & 8\textsuperscript{1}    & Finished & 0.34s (2 iter)                 \\
        ETHMAC                         & Timeout          & 6\textsuperscript{1}    & Finished & 156.55s (3 iter)               \\
        MOR1KX                         & Timeout          & 2\textsuperscript{1}    & Timeout  & Black-boxed                    \\
        WB\_RAM                        & Finished & 0.20s                           & Finished & 0.20s                          \\
        \midrule
        Subsystem 1                    & Timeout          & 298\textsuperscript{1}  & Finished & 0.13s (1 iter)                 \\
        Subsystem 2                    & Timeout          & 1012\textsuperscript{1} & Finished & 2.88s (1 iter)                 \\
        Subsystem 3                    & Timeout          & 2493\textsuperscript{1} & Timeout  & 34\textsuperscript{1} (5 iter) \\
        \bottomrule
        \multicolumn{5}{l}{\textsuperscript{1}\footnotesize{Number of properties that did not complete inside the time limit}}
      \end{tabular}
    \end{center}
  \end{minipage}
  \label{tab:comparison}
\end{table}%

We followed the approach presented in Algorithm~\ref{alg:semiformal_flow} and started the validation process verifying the IPs separately. In Phase 2, HWVerifyr verified successfully only the RAM memory IP, while all the other IPs needed Phase 3. In Phase 3, it verified the CAN IP after two iterations, the Ethernet IP after three iterations, and the processor IP was black boxed after five iterations. The first part of Table~\ref{tab:comparison} presents the results for Phases 2 and 3.

In Phase 4, the build-and-prove process started with the processor IP and the RAM memory IP. As mentioned above, it was not possible to verify the processor IP and, therefore, it was black boxed in Phases 4 and 5. HWVerifyr verified that subsystem in one iteration. Next, it added the CAN IP to the subsystem; however, it was not possible to verify it with the model checker. Therefore, it was necessary to switch to Phase 5. In Phase 5, HWVerifyr verified it in one iteration. Finally, it added the ETHMAC IP to the subsystem, but was unable to complete the verification process. The second part of Table~\ref{tab:comparison} presents the results for Phases 4 and 5.

Even though it was not possible to verify subsystem 3, SRA showed a considerable gain over the model checker alone. It is now possible to focus on the region left unproven, which is smaller than without SRA.

Our results show an improvement over JasperGold alone. Table~~\ref{tab:comparison} shows an improvement of 50\% in Phase 3 since it was possible to complete the formal proof for two more IPs than JasperGold. It also shows an improvement of 66\% in Phase 5, where our flow was able to complete the proof for two iterations in the ``build-and-prove'' system before timing out.

\subsection{SRA Validation}
We used the Ethernet and the CAN IPs to validate the SRA heuristic. As described in Section~\ref{sra}, SRA works with the registers the user has control over, i.e. configuration and control registers. For the validation process, we ran Phase 3 on the selected IPs.

The declaration of the configuration registers for the Ethernet IP is in the file ``eth\_registers.v''. SRA ranked them according to their COR and Table~\ref{tab:sra_ethmac} presents the results. The chosen time limit for all runs was 3600 seconds.

\begin{table}%
  \caption{Comparison of different ETHMAC registers for SRA validation}
  \begin{minipage}{\columnwidth}
    \begin{center}
      \begin{tabular}{lcccccc}
        \toprule
        Register                         & COR  & & Stopats & Time [s] & & Result \\
        \midrule
        MODER                            & 2880 & & 3       & Time-out & & 1178\textsuperscript{1} / 2\textsuperscript{2} \\
        MIICOMMAND                       & 381  & & 3       & Time-out & & 1171\textsuperscript{1} / 9\textsuperscript{2} \\
        CTRLMODER                        & 332  & & 1       & Time-out & & 1172\textsuperscript{1} / 6\textsuperscript{2} \\
        MIIMODER                         & 236  & & 2       & Time-out & & 1170\textsuperscript{1} / 8\textsuperscript{2} \\
        PACKETLEN                        & 214  & & 4       & Time-out & & 1172\textsuperscript{1} / 9\textsuperscript{2} \\
        \midrule
        MODER +\\MIICOMMAND               &      & & 6       & Time-out & & 1182\textsuperscript{1} / 2\textsuperscript{2} \\
        \midrule
        MODER +\\MIICOMMAND +\\CTRLMODER   &      & & 7       & 156.55s  & & 1186\textsuperscript{1} / 0\textsuperscript{2} \\
        \bottomrule
        \multicolumn{7}{l}{\textsuperscript{1}\footnotesize{Properties that either passed or failed}} \\
        \multicolumn{7}{l}{\textsuperscript{2}\footnotesize{Properties that did not complete inside the time limit}}
      \end{tabular}
    \end{center}
  \end{minipage}
  \label{tab:sra_ethmac}
\end{table}%

Table~\ref{tab:sra_ethmac} shows that SRA needed three iterations to find the minimum set of registers for the Ethernet IP, which are \textbf{MODER}, \textbf{MIICOMMAND}, and \textbf{CTRLMODER}. We ran the process with the other registers for comparison purposes. Furthermore, the low number of stop-ats for this set is a good indication that the system will not be over-constrained during the semiformal verification process.

The declaration of the configuration registers for the CAN IP is in the file ``can\_registers.v''. SRA ranked them according to their COR and Table~\ref{tab:sra_can} presents the results. The chosen time limit for all runs was 3600 seconds.

\begin{table}%
  \caption{Comparison of different CAN registers for SRA validation}
  \begin{minipage}{\columnwidth}
    \begin{center}
      \begin{tabular}{lccccccc}
        \toprule
        Register        & COR  & & Stopats & Time [s] & & Result \\
        \midrule                                       
        MODE            & 1177 & & 3       & Timeout  & & 169\textsuperscript{1} / 5\textsuperscript{2}  \\
        COMMAND         & 750  & & 4       & Timeout  & & 170\textsuperscript{1} / 8\textsuperscript{2}  \\
        CLOCK\_DIVIDER  & 705  & & 3       & Timeout  & & 165\textsuperscript{1} / 8\textsuperscript{2}  \\
        BUS\_TIMING1    & 312  & & 1       & Timeout  & & 150\textsuperscript{1} / 21\textsuperscript{2} \\
        BUS\_TIMING0    & 208  & & 1       & Timeout  & & 164\textsuperscript{1} / 7\textsuperscript{2}  \\
        \midrule                                       
        MODE +\\COMMAND  &      & & 7       & 0.34     & & 180\textsuperscript{1} / 0\textsuperscript{2}  \\
        \bottomrule
        \multicolumn{7}{l}{\textsuperscript{1}\footnotesize{Properties that either passed or failed}} \\
        \multicolumn{7}{l}{\textsuperscript{2}\footnotesize{Properties that did not complete inside the time limit}}
      \end{tabular}
    \end{center}
  \end{minipage}
  \label{tab:sra_can}
\end{table}%

Table~\ref{tab:sra_can} shows that SRA needed two iterations to find the minimum set of registers for the CAN IP, which are \textbf{MODE} and \textbf{COMMAND}. We ran the process with the other registers for comparison purposes. Again, the low number of stop-ats help to reduce the state space without over-constraining it.

\section{Conclusion}
\label{sec:conclusion}
We have presented our scalable hybrid verification approach for complex hardware systems. We described the advantages of the proposed methodology, which spans several steps in the hardware verification flow. The process begins with the formal verification of each IP and ends with the build-and-prove system that verifies incrementally bigger subsystems up to the complete architecture. The semiformal phases of the proposed methodology use the SRA heuristic to reduce the state space without over-constraining the architectures. Our results show that this methodology greatly benefits the verification flow of complex SoCs.

As future work, it should be possible to execute Phase 4 with different starting subsystems to create ``verified islands'' in the architecture when complete verification is not possible. It is also our goal to reduce the number of necessary stop-ats, to continue avoiding over-constraint. Finally, we want to add a smart time limit for the model checker since complex systems need more time to complete the verification task.


%
%

\bibliographystyle{IEEEtran} 
\bibliography{../bib/paper}

\begin{thebibliography}{10}
\providecommand{\url}[1]{#1}
\csname url@samestyle\endcsname
\providecommand{\newblock}{\relax}
\providecommand{\bibinfo}[2]{#2}
\providecommand{\BIBentrySTDinterwordspacing}{\spaceskip=0pt\relax}
\providecommand{\BIBentryALTinterwordstretchfactor}{4}
\providecommand{\BIBentryALTinterwordspacing}{\spaceskip=\fontdimen2\font plus
\BIBentryALTinterwordstretchfactor\fontdimen3\font minus
  \fontdimen4\font\relax}
\providecommand{\BIBforeignlanguage}[2]{{%
\expandafter\ifx\csname l@#1\endcsname\relax
\typeout{** WARNING: IEEEtran.bst: No hyphenation pattern has been}%
\typeout{** loaded for the language `#1'. Using the pattern for}%
\typeout{** the default language instead.}%
\else
\language=\csname l@#1\endcsname
\fi
#2}}
\providecommand{\BIBdecl}{\relax}
\BIBdecl

\bibitem{ITRS2015}
ITRS, ``{International Technology Roadmap for Semiconductors 2.0},'' Tech.
  Rep., 2015.

\bibitem{Turley2013}
\BIBentryALTinterwordspacing
J.~Turley, ``{Avoiding the SoC Verification Iceberg},'' 2013. [Online].
  Available: \url{https://www.eejournal.com/article/20130815-breker/}
\BIBentrySTDinterwordspacing

\bibitem{Foster2015}
H.~Foster, ``{Trends in functional verification},'' in \emph{Proceedings of the
  52nd Annual Design Automation Conference (DAC '15)}.\hskip 1em plus 0.5em
  minus 0.4em\relax New York, New York, USA: ACM Press, 2015, pp. 1--6.

\bibitem{Singhal}
\BIBentryALTinterwordspacing
V.~Singhal, ``{Oski Technologies-Formal Verification: Theory and Practice}.''
  [Online]. Available:
  \url{http://drona.csa.iisc.ernet.in/~deepakd/talks/formal-iisc-0306.pdf}
\BIBentrySTDinterwordspacing

\bibitem{Cimatti2013}
A.~Cimatti and et~al., ``{Software Model Checking SystemC},'' \emph{IEEE
  Transactions on Computer-Aided Design of Integrated Circuits and Systems},
  vol.~32, no.~5, pp. 774--787, 2013.

\bibitem{seligman2015formal}
E.~Seligman, T.~Schubert, and M.~V. A.~K. Kumar, \emph{{Formal verification: an
  essential toolkit for modern VLSI design}}.\hskip 1em plus 0.5em minus
  0.4em\relax Morgan Kaufmann, 2015.

\bibitem{Gajski2009}
D.~Gajski and et~al., \emph{{Embedded System Design}}.\hskip 1em plus 0.5em
  minus 0.4em\relax Boston, MA: Springer US, 2009.

\bibitem{Nenni2016}
D.~Nenni and D.~Dingee, \emph{{Prototypical - The Emergence of FPGA-Based
  Prototyping for SoC Design}}.\hskip 1em plus 0.5em minus 0.4em\relax
  CreateSpace Independent Publishing Platform, 2016.

\bibitem{Simm2015}
U.~Simm, S.~Rosenberg, E.~de~Kock, and P.~A. Hartmann, ``{Accellera Standards
  Technical Update},'' in \emph{2015 Design and Verification Conference and
  Exhibition}, 2015.

\bibitem{EDA-013}
H.~Foster, ``{Applied Assertion-Based Verification: An Industry Perspective},''
  \emph{Foundations and Trends{\textregistered} in Electronic Design
  Automation}, vol.~3, no.~1, pp. 1--95, 2009.

\bibitem{Mukherjee2015}
R.~Mukherjee and et~al., ``{Hardware Verification Using Software Analyzers},''
  in \emph{2015 IEEE Computer Society Annual Symposium on VLSI}.\hskip 1em plus
  0.5em minus 0.4em\relax IEEE, jul 2015, pp. 7--12.

\bibitem{Herber2014}
P.~Herber, ``{The RESCUE Approach - Towards Compositional Hardware/Software
  Co-verification},'' in \emph{2014 IEEE Intl Conf on High Performance
  Computing and Communications (HPCC)}.\hskip 1em plus 0.5em minus 0.4em\relax
  IEEE, aug 2014, pp. 721--724.

\bibitem{Groe2006}
D.~Gro{\ss}e and et~al., ``{HW/SW co-verification of embedded systems using
  bounded model checking},'' in \emph{Proceedings of the 16th ACM Great Lakes
  symposium on VLSI - GLSVLSI '06}.\hskip 1em plus 0.5em minus 0.4em\relax New
  York, New York, USA: ACM Press, 2006, p.~43.

\end{thebibliography}

\end{document}